# E-Beam Induced Micropattern Generation and Amorphization of L-Cysteine-Functionalized Graphene Oxide Nano-composites

Y. Melikyan[1], H. Gharagulyan[1, 2, *], A. Vasil'ev[1], V. Hayrapetyan[1], M. Zhezhu[1], A. Simonyan[1], D. A. Ghazaryan[3, 4], M. S. Torosyan[1], A. Kharatyan[1], J. Michalicka[5], M. Yeranosyan[1, 2]

[1]Innovation Center for Nanoscience and Technologies, A.B. Nalbandyan Institute of Chemical Physics NAS RA, 5/2 P. Sevak str., Yerevan 0014, Armenia
[2]Institute of Physics, Yerevan State University, 1 A. Manoogian, Yerevan 0025, Armenia
[3]Moscow Center for Advanced Studies, Kulakova str. 20, Moscow, 123592, Russia
[4]Laboratory of Advanced Functional Materials, Yerevan State University, 1 A. Manoogian, Yerevan 0025, Armenia
[5]CEITEC - Brno University of Technology, Purkyňova 123, 612 00, Brno, Czech Republic
*Author to whom correspondence should be addressed: herminegharagulyan@ysu.am

**Abstract**

The evolution of dynamic processes in graphene-family materials are of great interest for both scientific purposes and technical applications. Scanning electron microscopy and transmission electron microscopy outstand among the techniques that allow both observing and controlling such dynamic processes in real time. On the other hand, functionalized graphene oxide emerges as a favorable candidate from graphene-family materials for such an investigation due to its distinctive properties, that encompass a large surface area, robust thermal stability, and noteworthy electrical and mechanical properties after its reduction. Here, we report on studies of surface structure and adsorption dynamics of L-Cysteine on electrochemically exfoliated graphene oxide's basal plane. We show that electron beam irradiation prompts an amorphization of functionalized graphene oxide along with the formation of micropatterns of controlled geometry composed of L-Cysteine-Graphene oxide nanostructures. The controlled growth and predetermined arrangement of micropatterns as well as controlled structure disorder induced by e beam amorphization, in its turn potentially offering tailored properties and functionalities paving the way for potential applications in nanotechnology, sensor development, and surface engineering. Our findings demonstrate that graphene oxide can cover L-Cysteine in such a way to provide a control on the positioning of emerging microstructures about 10-20 μm in diameter. Besides, Raman and SAED measurement analyses yield above 50% amorphization in a material. The results of our studies demonstrate that such a technique enables the direct creation of micropatterns of L-Cysteine-Graphene oxide eliminating the need for complicated mask patterning procedures.

**Introduction**

Graphene oxide (GO) is a graphene's derivative decorated with oxygen functional groups on its basal plane [1, 2]. Due to the large surface area and its structure with plentiful oxygen functional groups, as well as both hydrophilic and hydrophobic parts, GO exhibits significant potential in applications, such as in biosensors [3], energy storage [4], catalysis [5] *etc.* The area of potential applications and physicochemical properties of this material can be drastically increased and improved *via* its functionalization [6]. From this point of view, amino acid (AA)-functionalized GO plays a key role in today's biomedicine, namely, in drug delivery, tissue engineering and sensing technologies due to its relatively long-term biocompatibility and functionality [7, 8]. Functional groups of AAs and GO enhance the material's interaction with biological systems, and thus, increase its functional capabilities [9].

L-Cysteine (Cys), as a polar AA has garnered significant attention from the scientific community owing to the advancement of GO modification [10, 11]. In [12], a thick Cys functionalized GO was fabricated under atmospheric pressure and effective suppression of the GO sheets' aggregation was shown. Cys acts as a reduction agent for GO converting it into reduced GO (rGO), which possesses similar properties as graphene [13], but with more defects and, thus, less quality [14, 15]. The applications of such structures are further expanded due to their catalytic properties [16-18]. Notably, when exposed to an e-beam irradiation, those exhibit interesting behavior, such as reduction, crosslinking, polymerization, radical formation, functional group modification *etc.*

The e-beam irradiation technology shows promising potential for the extensive creation of rGO while maintaining accurate control over its oxygen portions [19, 20]. From this point of view, scanning electron microscopy (SEM) and transmission electron microscopy (TEM) allow for real-time observation and analysis of dynamic processes occurring at the nanoscale under the influence of an e-beam. They give the opportunity not only to study the morphology of these systems at an atomic scale, but also, to influence the matter during the observation. One of the pioneering observations of intensive interactions between e-beam and matter was presented in [21]. It led to alterations in the structure and properties of the material. Since then, significant progress has been made in the field, with numerous studies conducted. For example, in [22], the authors explore the utilization of e-beam cross-linked functional polymers for precise positioning of multiple proteins at the nanoscale. On the other hand, e-beam irradiation can lead to several degradation processes, such as deamination, decarboxylation, and desulfurization [2323]. In [24], the authors show that for different e-beam energies and electron doses, GO layers undergo a phase transition to an amorphous carbon. Notably, amorphous structural models for GO were proposed in [25]. Here, the authors provided some estimates on geometric structures, thermodynamic stabilities, and electron density of states of amorphous GO models. In addition, the reduction of thermal conductivity and its tuning, namely, structural deformations and amorphous limit of GO is theoretically discussed in [26]. It is worth noting, that similar phenomenon is observed under ion-beam irradiation [27]. Here, the disorder parameter of GO derived from Raman spectra exhibited an elevation indicating formation of defects within its lattice, which ultimately lead to amorphization. Furthermore, in [28], the tuning possibilities of GO properties are discussed by ion beam, namely, modification of the ratio of $sp^2$ and $sp^3$ hybridization. The impact of gamma irradiation at various doses on a water dispersion of graphene quantum dots mixed with L-Cysteine and IPA was explored in [29]. It was revealed that after gamma irradiation not only structural modifications but also changes in optical properties were observed.

In general, the fabrication of organized arrays of nanomaterials by a variety of modern techniques, such as nanografting, conductive atomic force or scanning probe microscopies, dip-pen nanolithography, nanoimprint lithography, e-beam or photo- lithography, focused ion beam, polymer/DNA assisted templating and pulsed laser deposition *etc.*, is still a challenge since it is either expensive or time-consuming [30, 31]. Notably, all these techniques can be templCte- replica, induced, and free providing different types of nanopatterning, such as chemical patterning, topographical patterning, 3D patterning, combinatorial patterning, and nano-biopatterning *etc.* However, high resolution patterning, namely, the generation of evenly distributed submicron sized structures on flat surfaces by fast and simple fabrication techniques is yet of great interest. In this study, we propose a straightforward method to generate micropatterns of Cys-GO nanocomposites with varying shapes, sizes, inter-features, and controlled positions through e-beam irradiation. Furthermore, we also report on the amorphization of our nanocomposites boosted by the e-beam irradiation. Our results demonstrate that such technique enables the direct creation of micropatterns of Cys-GO eliminating the need for complex mask patterning procedures.

1. Materials and Methods

**Materials.** The graphite foil, solvents, L-Cysteine, and all the chemicals used in the experiments were purchased from Sigma-Aldrich Chemical Co.

**GO Synthesis.** The electrochemical exfoliation of graphite foil was conducted in a three-electrode cell of Zahner photo-electrochemical station, where a platinum electrode served as cathode, graphite as anode, and a silver chloride electrode (SCE) as reference. The distance between the counter-electrodes was 1.5-2 cm, and the electrolyte volume was 75 mL. A 98 % $H_2SO_4$ was used as intercalation electrolyte. Intercalation was performed in a galvanostatic mode at a specific current of 50 mA. During this process, the applied potential did not exceed 2.2 V. The duration of the process was 20 min. The exfoliation was carried out in 0.1 M $(NH_4)_2SO_4$ for potentiostatic mode at 5 V and lasted 30 min. After completing the process, expanded graphite flakes, containing GO were collected and washed by filtration. Then, obtained flakes were ultrasonicated for 30 min to liberate GO from large graphite flakes. Finally, brownish GO solution was separated from large non-dispersible graphite flakes by centrifugation. The resulting transparent brown solution was used for further experiments.

**L-Cysteine Functionalization of GO.** For preparation of Cys-GO, 100 mL of GO (concentration ~ 50 mg/L) was added in a flask and the solution was stirred at room temperature for 30 min, and then, sonicated for another 30 min. Then, 80 mg of Cys was added into 80 mL of distilled water and properly stirred. The Cys solution was added to the GO solution and the reaction mixture was ultrasonicated at the room temperature for 1 h. The obtained solution was drop-casted on Si substrate and dried at 80° C for 30 min.

1.1. Characterization

Crystallographic data for the GO powder and Cys-GO nanocomposites were obtained by XRD analysis (MiniFlex, Rigaku). Chemical compound analysis of mentioned materials was conducted using FTIR-ATR spectrometry (Spectrum Two, PerkinElmer). The bond types and hybridizations of GO were identified using Raman spectroscopy (XploRA PLUS, HORIBA). The O/C ratio, chemical composition of GO, and Cys-GO nanocomposites were determined by HR XPS analysis (KRATOS, Axis Supra[+], Shimadzu). To monitor the mass loss over time, the GO and Cys-GO samples were subjected to thermal gravimetric analysis (TGA 8000, PerkinElmer). Optical characterization of these materials was performed using PL spectroscopy (Cary Eclipse Fluorescence Spectrometer, Agilent). The absorbing properties of dispersed GO and Cys-GO nanocomposites were investigated using UV-Vis spectrophotometry (Cary 60, Agilent). The morphology and chemical composition of GO and Cys-GO were analyzed *via* a scanning electron microscope (SEM) Prisma E (Thermo Fisher Scientific) and an image corrected transmission electron microscope (TEM) Titan 60-300 Themis (ThermoFisher Scientific) operated at 300 kV and equipped by high-annular angular dark-field detector for scanning TEM imaging (STEM-HAADF) and Super-X detector for energy dispersive X-ray spectroscopy elemental mapping in STEM mode (STEM-EDX). Particle size analysis and zeta potential's value of the synthesized and functionalized materials were implemented utilizing the dynamic light scattering (DLS) technique (Litesizer 500, Anton Paar).

## 3. Results and Discussion
### 3.1. Structural Analysis of GO

The characteristic spectra of electrochemically exfoliated GO are presented in Fig.1. The analysis of all the measurements showed that the synthesized material corresponds to the known ones in the literature [32]. According to Fig.1(a), the UV-Vis spectrum of GO exhibits characteristic peaks at around 248 nm, which is attributed to π–π* transitions from the aromatic C–C bonds, and at around 305 nm, which is attributed to n-π* transitions from C-O bonds. Fig.1(b) shows the photoluminescence (PL) spectrum of the aqueous solution of GO. The PL peaks were observed at 401 nm correspond to π–π* transitions, and at 535 nm to n-π* transitions in case of the excitation wavelength of 375 nm. The XRD spectrum of synthesized GO powder is shown in Fig.1(c). The diffraction peak of electrochemically exfoliated GO is observed at 11°, which corresponds to the 001 plane. The Raman spectrum of our GO shown in Fig.1(d) exhibits two characteristic bands located at wavenumbers of 1347.3 $cm^{-1}$ and 1595.9 $cm^{-1}$ that correspond to the scattering from local defects and disorder present in carbon (*D* band), and in-plane tangential stretching of the C−C bonds in the graphitic structure (*G* band). Notably, the obtained $I_D/I_G$ ratio is 0.852. A less intense *2D* band for GO samples is seen at 2690 $cm^{-1}$, which indicates its reduction. Chemical bonding characteristics of synthesized material was revealed by the FTIR-ATR spectra as it is shown in Fig.1(e). The bands at the wavenumber of about 1700 $cm^{-1}$, 1585 $cm^{-1}$, 1448 $cm^{-1}$, 1131 $cm^{-1}$ can be assigned to the C = O carbonyl, C = C, C-OH carboxylic, C-O-C epoxy group stretchings, respectively [33]. The thermal stability of the exfoliated GO was evaluated through thermal gravimetric (TGA) analysis. As shown in Fig.1(f), it decomposes in three steps: at the first stage, the mass loss occurs at 40° C due to the loss of water molecules; at the second stage, it occurs at 170° C up to 270° C due to the loss of oxygen containing groups, at the third stage, a total decomposition of GO was observed at 400° C. In Fig.1(g), the zeta potential of GO is depicted, for the neutral pH, it is about -33.28 mV. The mean hydrodynamic diameter measured is approximately 850 nm for the synthesized GO. The SEM image of GO's morphology is shown in Fig.1(i). It appears as randomly distributed thin sheets with irregular edges, wrinkled and crumpled surfaces because of their scrolling and folding.

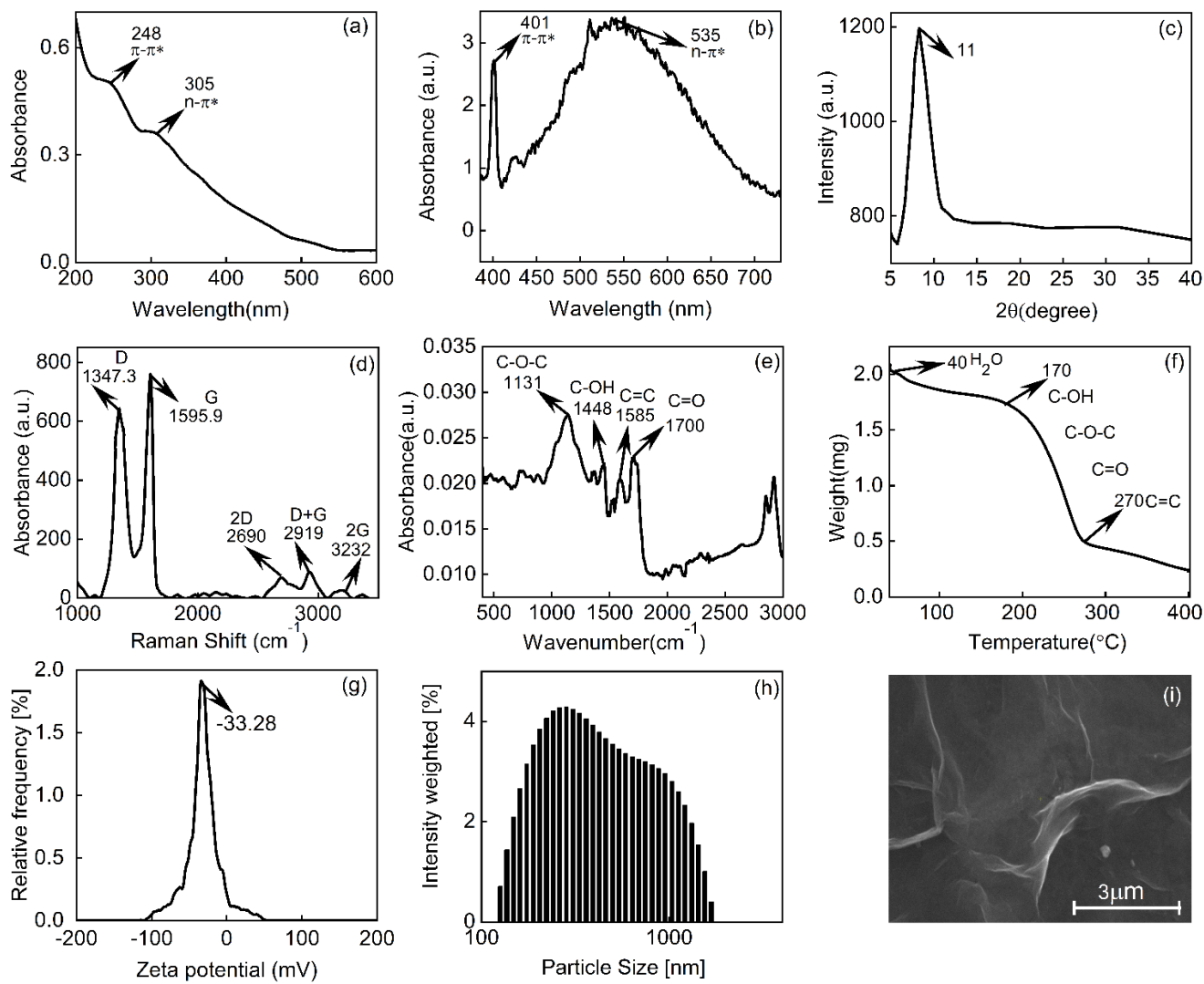

Fig.1 Characterization of the synthesized GO: (a) UV-Vis spectrum; (b) PL emission spectrum (the excitation wavelength is 375nm); (c) XRD spectrum; (d) Raman spectrum; (e) FTIR-ATR spectrum; (f) TGA analysis (heating was performed at 5°C per minute at a flow rate of 60 mL/min); (g) Zeta potential; (h) Particle size distribution (the system was equipped with 658 nm laser diode as a light source and measured the particle size at a scattering angle of 175°); (i) SEM image. All measurements were conducted at an elevated temperature of 25° C.

### 3.2. Structural Analysis of Cys-functionalized GO

The optical properties of pristine and Cys-functionalized GO were determined by UV–Vis spectroscopy and the obtained spectra are presented in Fig.2(a). As one can see, the samples show typical absorption peaks in the UV region. In the case of Cys-modified GO, the peak at around 305 nm corresponding to n-π* transitions from C-O bond showed a blueshift by 8 nm. The other absorption band at around 248 nm that is attributed to π–π* transitions from the aromatic C–C bond was not modified. In Fig.2(b), PL spectra of pristine and Cys-modified GO are presented. Notably, in comparison to the PL peaks of GO, for Cys-functionalized GO, the intensity of PL emission peak corresponding to π–π* transitions is decreased and the peak attributed to n-π* transitions is quenched. XRD spectra of the Cys-functionalized GO and pristine GO are shown in Fig.2(c). In comparison with pristine GO, which has a diffraction peak at 11°, the peak of Cys-GO is shifted to the higher angles having a maximum at 11.3°. This is due to the changes in the interplanar distances between layers as a result of the interaction between GO and Cys. Also, the corresponding peak undergoes significant changes in the intensity due to the degradation of crystalline structure. Besides, the second peak at 25° was observed, which corresponds to the rGO. To identify functional groups on the surface of Cys-functionalized GO, FTIR-ATR analysis was employed (see Fig.2(d)). The spectrum of Cys-GO clearly demonstrates successful functionalization evidenced by the presence of certain bands not observed in the GO spectra. Particularly, for Cys-GO nanocomposites, the band at 2895 cm$^{-1}$ shows the vibrations of the -NH$_2$ bonds, and the -SH bonds were observed at 2541 cm$^{-1}$. The bands at 1492 cm$^{-1}$ and 1733.2 cm$^{-1}$ correspond to the symmetrical and asymmetrical bendings of the -NH$_2$ group of Cys [34]. The bands at 1062 cm$^{-1}$ and 1139.5 cm$^{-1}$ are attributed to the -CO and -CN groups. Besides, the bands at 1492 cm$^{-1}$ and 1405 cm$^{-1}$ refer to asymmetrical and symmetrical vibrations of the -COO group. In addition, the bands at 666.8 cm$^{-1}$ and 786.4 cm$^{-1}$ correspond to the -CS stretching vibrations. Finally, the band at 510 cm$^{-1}$ corresponds to the S-S vibration. Thus, Cys can interact with GO due to hydroxyl and carboxyl groups of GO through S-S, NH$_2$, COO and OH bonds, which were observed in the ATR spectra. The TEM image of Cys-GO morphology is shown in Fig.2(e). The decoration of GO surface with Cys has a negligible impact on its wrinkled surface, which is originated from the layering of the sheets. On the other hand, the picture clearly shows the crystalline structure of Cys. Zeta potential of the above-mentioned structures was also measured, as it can be seen from Fig.2(f), for the neutral pH, it is about -33.28 mV in case of pristine GO, and -6.77 mV for Cys-GO nanocomposites. The decrease of zeta potential might be attributed to the adsorption of L-cysteine on the surface of GO structure, mainly to the reduction of oxygen-containing functional groups [35]. The obtained comparison for the number-weighted average diameter distribution function for the particle-sizes of GO and Cys-GO are presented in Fig.2(g). The average hydrodynamic diameter was approximately 850 nm for GO, and, in case of Cys-GO nanocomposites, it increased to 2125 nm. The thermal stability of the synthesized pure GO and Cys-GO was examined by thermal gravimetric analysis (see Fig.2(h)). Unlike pristine GO, for Cys-GO nanocomposites, decomposition was observed again in three steps, but it was less pronounced. The weight loss at 29° C was observed due to the presence of moisture in the sample. Similarly, the weight decrease was observed at 75° C and at 294° C for the oxidative groups and final decomposition, respectively. Therefore, the functionalization of GO with Cys leads to the decrease of its thermal stability.

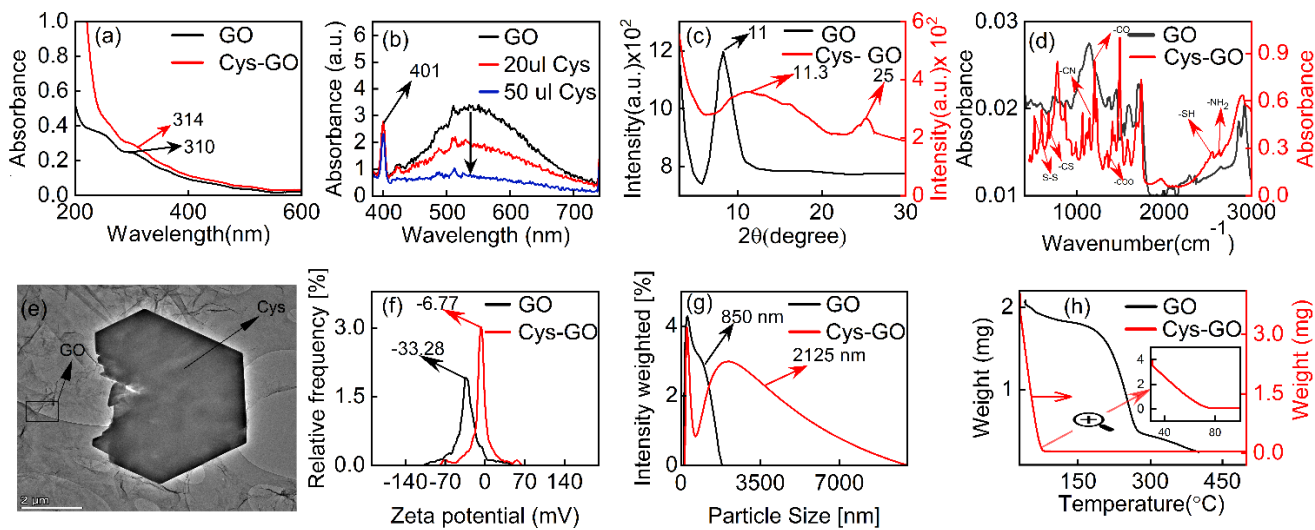

Fig.2 Characterization of Cys-GO nanocomposites and its comparison to GO. (a) UV-Vis spectra of GO (black) and Cys-GO (red); (b) PL emission spectra of GO (black) and Cys-GO (blue and red); (c) XRD analysis of GO (black) and Cys-GO (red); (d) FTIR-ATR spectra of GO (black), Cys-GO (red); (e) TEM image of Cys-GO morphology; (f) Zeta potential of GO (black) and Cys-GO (red); (g) Particle size distribution of GO (black) and Cys-GO (red); (h) TGA analysis of GO (black), Cys-GO (red).

XPS studies of GO and Cys-functionalized GO were performed using HR XPS Axis Supra+. The XPS spectral analysis was performed by CasaXPS software. At first, a fast screening was done for the obtaining of a wide spectrum scan XPS spectra (see Fig.3(a)). Here, we obtain that the O/C ratio is 0.4, it was calculated according to [36]. Our results for Cys-GO show the S2p peak at 164.11 eV and 165.22 eV for $S2p_{3/2}$ and $S2p_{1/2}$ high spin and low spin orbitals, respectively. This confirms surface functionalization of GO by Cys. Notably, this peak was not observed in the case of pristine GO, which means that our synthesized material is sulfate free. In XPS wide spectrum, the Si is denoted by the asterisk, which corresponds to the substrate. Fig.3(b)-(e) present fitted and original spectra for Cys-GO nanocomposite's elements, namely, C1s, O1s, S2p and N1s, respectively. The presence of S2p and N1s imply for the successful functionalization of GO by Cys.

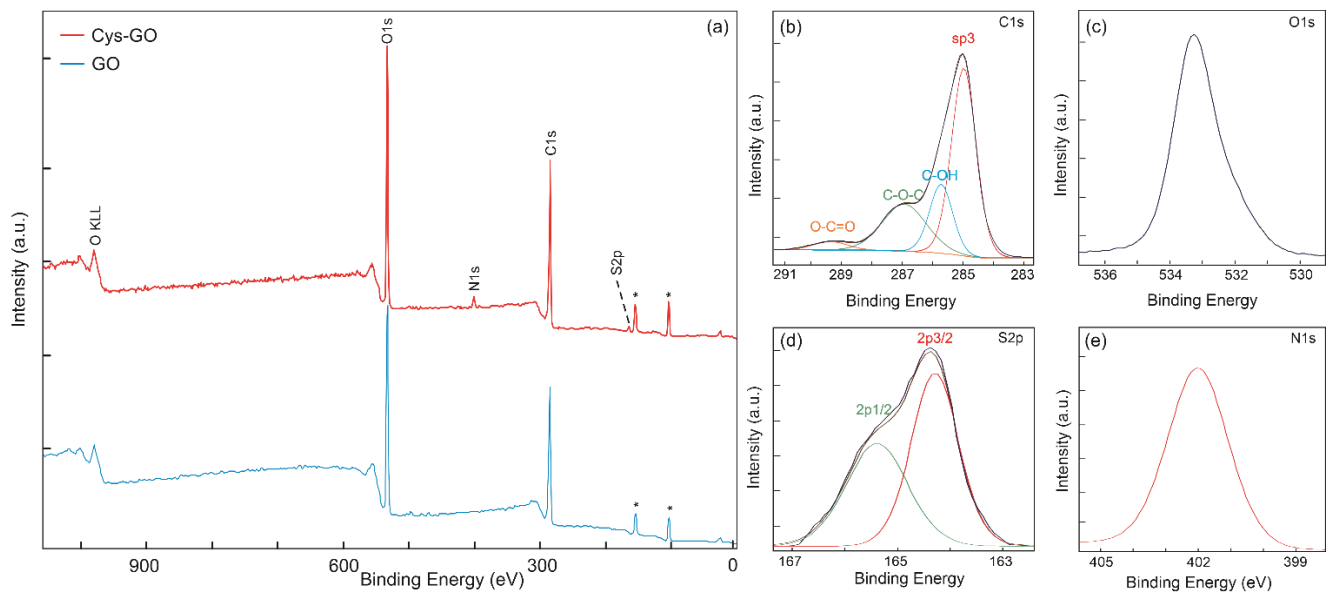

Fig. 3 XPS spectra of Cys-GO and its comparison to GO: (a) XPS survey scan spectra for GO (blue) and Cys-GO (red); and high resolution XPS spectra for Cys-GO: (b) C1s peak, (c) O1s peak, (d) S2p peak and (e) N1s peak. Binding energies were referenced to the C1s main peak of carbon tape at 285 eV.

Table 1 represents XPS binding energy peak positions for GO and Cys-GO.

Table 1

| | Cys-GO | | | | | GO | |
|---|---|---|---|---|---|---|---|
| | *Peak position (eV)* | | | | | *Peak position (eV)* | |
| C1s | sp³C | 284.97 | N1s | 402 | C1s | sp³C | 285.02 |
| | C-O | 285.72 | $S2p_{3/2}$ | 164.11 | | C-O | 286.85 |
| | C=O | 286.93 | $S2p_{1/2}$ | 165.22 | | O-C=O | 289.16 |
| | O-C=O | 289.35 | | | | | |
| O1s | | 533.27 | | | O1s | | 532.91 |

### 3.3. E-Beam Irradiation Analysis of Cys-functionalized GO

The surface morphology changes and chemical composition for Cys-functionalized GO were investigated by TEM operated in scanning (STEM) mode with the following observation conditions: accelerating voltage: 300 kV;

beam current: 700 pA; scan frame: 512x512; pixel size: 21.6 nm; dwell time: 4 µs; illuminated area: 23.2 µm$^2$. Fig.4(a) shows the morphology of Cys-GO nanocomposites before e-beam irradiation, and Fig.4(b) presents the result of the dynamic processes observed during e-beam irradiation. SEM images of pristine Cys before (a), and after (b) the e-beam influence are shown in Supplementary Materials (see Fig.S1). As it is obvious from the Fig. 4(a) and (b), under the influence of e-beam, structural changes are observed. The Raman spectra, presented in Fig. 4 (c), also confirm the proposed hypothesis of amorphization similar to [37, 38]. It is worth to mention that Raman analysis were done at the irradiated part of the sample and e-beam irradiation was performed by SEM. Here, $I_D/I_G$ ratio for Cys-GO is increased from 1.2 to 1.8 after the e-beam irradiation. Besides *D''* band between *D* and *G* bands at approximately 1506 cm$^{-1}$ corresponding to interstitial defects associated with amorphous sp$_2$-bonds expansion, appears shifted to 1522 cm$^{-1}$. Additionally, *D'* band's intensity increase also indicates progressive material's amorphization. The $I_{D''}/I_G$ ratio represents an increase in the crystallinity or decrease in the amorphicity. In our case, it was 0.46 before e-beam irradiation and became 0.48 after it. On the other hand, the band corresponding to the S-S bonds also undergoes a great change after e-beam exposure, namely, from 497 cm$^{-1}$ to 509 cm$^{-1}$ due to disulfide bond distortion and sulfhydryl bond formation [39]. Fig.4(d) and (e) show the same process (before and after e-beam irradiation) observed for relatively large part of the sample. The observations were done for the following SEM parameters: accelerating voltage: 5, 10, 20 kV; beam current: 100 pA, 0.6 nA, 1.2 nA, 4.5 nA; spot size: 3.5, 4.5, 5.0, 6.0; scan frame: 768x512; pixel size: 0.4 Mpx; dwell time: 4 µs. From Fig.4(e) one can see that it is showing multiple spherical micropatterns of Cys-GO nanocomposites which occur on GO template. The micropatterns' architecture depends not only on the e-beam irradiation parameters, but also on the Cys-functionalization process of GO, namely the pH, concentration and temperature of the mixture. Notably, the amorphization process can be confirmed in-situ in TEM by selected area electron diffraction (SAED) pattern acquisition. The SAED acquired before the TEM irradiation with a defocused e-beam revealed a well-defined diffraction pattern (see Fig.4(f)), attesting the material's crystallinity, which becomes amorphous during the e-beam exposure for 15 min (see Video in Supplementary Materials). The resulting SAED patterns during and after the amorphisation are shown in Fig.4(g)-(i). The EDX spectra (Fig.S2(a) in Supplementary Materials) obtained from the sample showed characteristic energy lines corresponding to C-K, N-K, S-K and O-K signal, while the elemental maps showed a distribution of the above-mentioned elements across the material surface (see Fig.S2(b)). The obtained EDX results showed that GO has C and O in their structure, while Cys-functionalized GO sample was mainly composed of C, O and S and N. The atomic percentages for Cys-modified GO are: C 41.5% at%; O 4.98% at%, N 11.62% at%; S 86.3% at%, respectively.

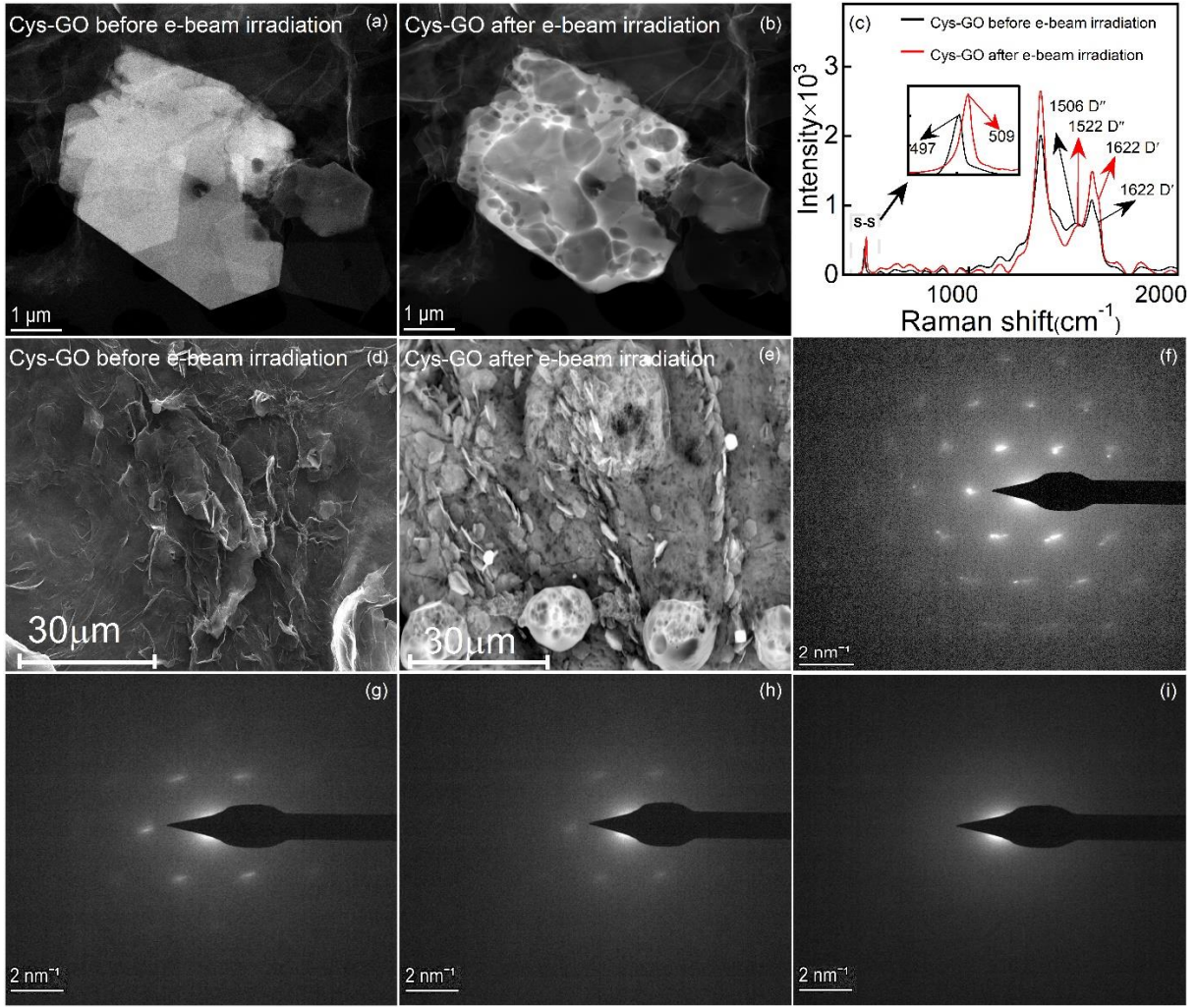

Fig. 4. STEM and SEM Imaging of Cys-GO degradation: STEM-HAADF imaging of (a) Cys-GO before e-beam exposure and (b) after 15 min e-beam exposure; (c) Raman spectra of Cys-GO structure before and after e-beam irradiation; (d) SEM image of Cys-GO structure from relatively large part of the sample before e-beam irradiation; (e) created Cys-GO micropatterns after e-beam irradiation (f) SAED diffraction pattern acquired before the 15 min irradiation in TEM mode; (g) SAED diffraction pattern at the beginning; (h) in the middle; and (i) end of 15 min irradiation process, respectively.

The e-beam interacts with the Cys-GO intensively, changing its structure and properties with the varying dose. For 300 kV accelerating voltage, the dynamic process observed in the Cys-GO takes longer time than for 20 kV. However, it is worth to mention that besides the amorphization due to the bond breaking, evaporation from the edges is also possible for the structure due to the OH groups of Cys and GO. For the mentioned parameters from the equation (1), the irradiation dose can be estimated as [23]:

$$a = \frac{(F \times E \times t)}{S}, \tag{1}$$

where $F$ is the flux of the electrons, $E$ is the electron energy, $t$ is irradiation time, $S$ – the area. And Fluence can be estimated as [40]:

$$Fluence = \frac{I \times t}{e \times S}, \tag{2}$$

where $I$ is the current, $t$ - time, e is elementary charge, $S$ – the area. The estimated irradiation dose for our sample is 2.72 A·s/cm$^2$, the fluence is 1.7·10$^{19}$ electron/cm$^{-2}$. A comparative assessment of the absorbed irradiation doses by different systems is given in Table 1 (see in Supplementary Materials).

It is worth noting, for pristine Cys, the dynamic process was observed in the form of blooming flower (see Fig.S1(b) in Supplementary Materials) without controlling the process, while for the Cys-GO due to the GO layer enveloping the Cys, a clear micropattern formation with controlled parameters was observed (see a single micropattern's image in Fig.S3 of Supplementary Materials). These micropatterns' properties depend on the penetration depth of electrons at different accelerating voltages: at a higher density of electrons strikethrough, it results in a more efficient energy transfers to a Cys-GO surface. Our findings show that for different values of the accelerating voltage extraordinary phenomena are observed. It was shown that the described dynamic processes were completely of different nature: beating at 10 kV, opening and closing and blooming at 20 kV, and polymerizing at 300 kV. It is worth nothing that created micropatterns can be distributed more uniformly in case of precise centrifugation-induced size-fractionation. For now, we don't have enough understanding to explain the observed phenomena and there is still a need for further studies. However, similar effects were observed for Cys-functionalized MoS$_2$ as well (see Fig.S4 in Supplementary Materials).

**Conclusion**

In summary, GO synthesis through electrochemical exfoliation and its functionalization by Cys was done in this work. The morphology dynamics, and eventually, amorphization of Cys-GO induced by the e-beam irradiation were studied. Comprehensive XPS analysis revealed the picture of elemental composition of such samples as well as the chemical and electronic states of the atoms. SEM and STEM manipulation techniques allowing precise control over micrometer-scale positioning of the Cys-GO nanocomposites were suggested. Here, detailed analysis at various doses, currents, and accelerating voltages made it possible to create Cys-GO micropatterns with controlled shapes, sizes, inter-feature spaces, and position. The gained results provide a new pathway for the non-lithographic fabrication process of the Cys-GO and similar structures. We believe, that our work will inspire future research on novel graphene-family materials in diverse research areas, such as adsorbents with high selectivity and good regenerability, membranes for nanofiltration, aerogels, photosensitive and purification agent *etc.*

**Supplementary Materials**

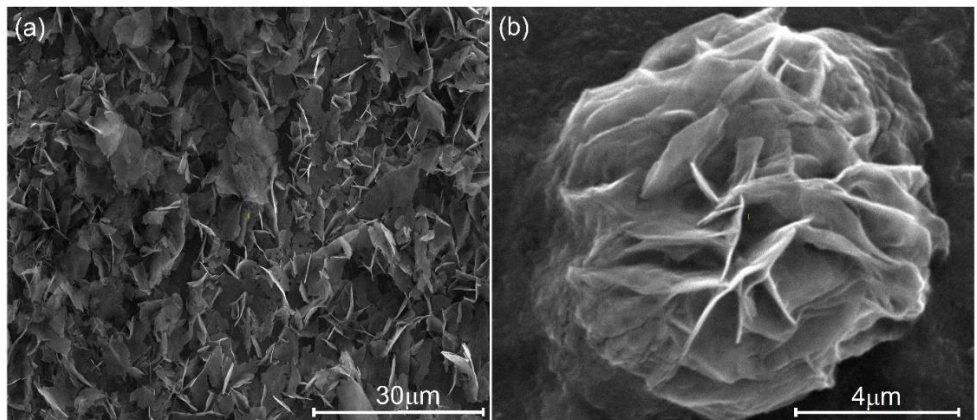

Fig.S1. SEM images of pristine Cys (a) before and (b) after the e-beam influence.

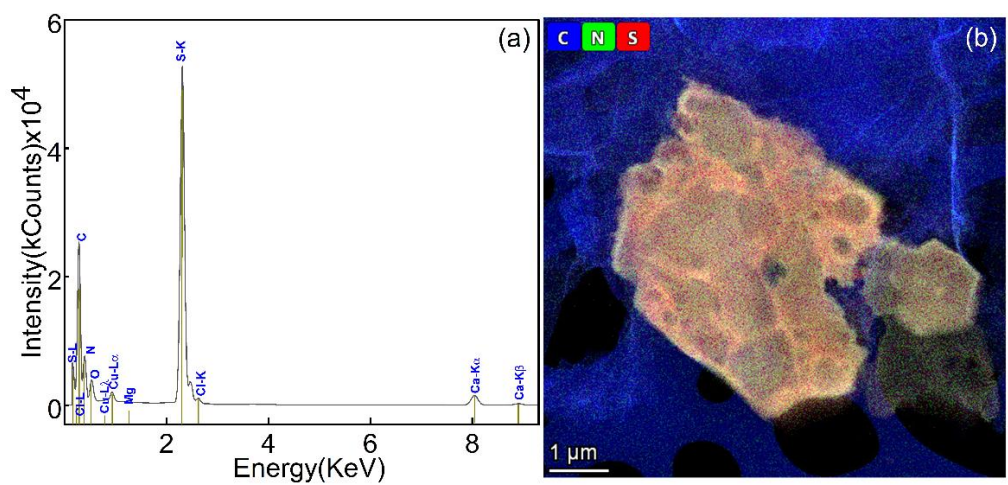

Fig.S2. EDX analysis: (a) analysis of chemical elements in the material from collected EDX spectrum, (b) an overlap of STEM-EDX chemical composition maps of C-K, N-K and S-K net intensities.

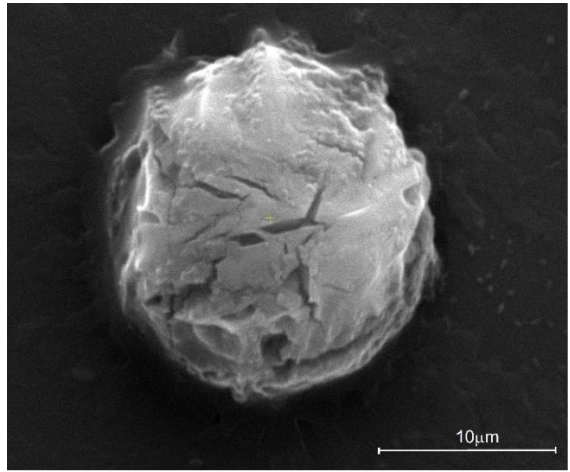
Fig.S3. SEM image of an isolated Cys-GO micropattern after e-beam irradiation.

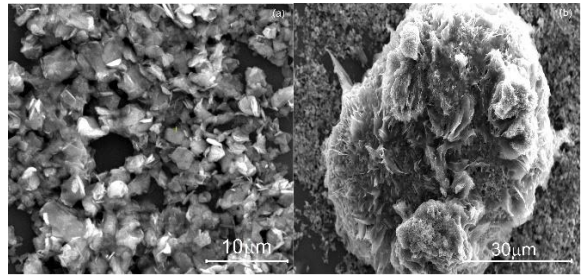
Fig.S4. SEM images of Cys-MoS$_2$ structure: (a) before and (b) after e-beam irradiation.

**Table 1**. Comparison of absorbed irradiation doses by different systems.

| Ref | System | Irradiation sources and Conditions | Absorbed irradiation dose and Fluence |
|---|---|---|---|
| [1] | GO suspension in isopropanol (volume ratio 3:2) | e-beam under room temperature 1.8 MeV 5 mA | 0,5,15,20,40 kGy |
| [2] | GO | Carbon ion beam 80 MeV 1 pnA | $1\times10^{11}$, $1\times10^{12}$, $1\times10^{13}$ ions/cm$^2$ |
| [3] | GO | Au ions irradiation 120 MeV 0.5 pnA | from $10^{10}$ to $10^{13}$ ions/cm$^2$ |
| [4] | GO papers | Ti ion implantation accelerating voltage 40kV 6mA | $5\times10^{16}$, $8\times10^{16}$, $1\times10^{17}$, $3\times10^{17}$ ions/cm$^2$ |
| [5] | GO films | He and Ga ions irradiation<br>500 keV<br>10 nA/cm$^2$ | from $1\times10^{14}$ to $5\times10^{16}$ ions/cm$^2$ |
| [6] | Graphene quantum dots | Gamma irradiation<br>using Co-60 as an irradiation source | 25, 50 and 200 kGy |
| [7] | Graphite oxide | e-beam<br>at room temperature, under ambient air conditions | 50~360 kGy |
| [8] | GO | e-beam<br>5 keV<br>~250 pA | from $7.5 \times 10^{17}$ to $1.5 \times 10^{19}$ electrons/cm$^2$ |
| [9] | GO thin films | e-beam<br>25 keV<br>irradiation time 1 min, 33 min, 100 min. | $3\times10^{11}$ electrons/cm$^2$, $1\times10^{13}$ electrons/cm$^2$ and $3\times10^{13}$ electrons/cm$^2$ |
| [10] | L-Cysteine | e-beam<br>800eV<br>e-beam diameter 3 mm<br>irradiation time 3 min<br>~ 1 μA | 250 MGy |
| [11] | GO | e-beam<br>accelerating voltage 3 and 10keV<br>150pA-40mA | 30 to 3000 mA·s/cm$^2$ |
| Our work | L-Cysteine-GO | e-beam<br>300kV<br>700pA<br>irradiation time 15min | 2.72 A·s/cm$^2$<br>Fluence: $1.7\times 10^{19}$ electrons/cm$^2$ |